# Optimal hierarchical modular topologies for producing limited sustained activation of neural networks

Marcus Kaiser[1,2,3*] and Claus C. Hilgetag[4,5]

1 School of Computing Science, Newcastle University, United Kingdom
2 Institute of Neuroscience, Newcastle University, United Kingdom
3 Department of Brain and Cognitive Sciences, Seoul National University, Korea
4 School of Engineering and Science, Jacobs University Bremen, Germany
5 Department of Health Sciences, Boston University, USA

**Correspondence:**
Dr Marcus Kaiser
School of Computing Science
Newcastle University
Claremont Tower,
Newcastle upon Tyne, NE1 7RU, United Kingdom
m.kaiser@ncl.ac.uk

**Running title:**
Optimal hierarchical networks for functional criticality

**Abstract**
An essential requirement for the representation of functional patterns in complex neural networks, such as the mammalian cerebral cortex, is the existence of stable regimes of network activation, typically arising from a limited parameter range. In this range of limited sustained activity (LSA), the activity of neural populations in the network persists between the extremes of either quickly dying out or activating the whole network. Hierarchical modular networks were previously found to show a wider parameter range for LSA than random or small-world networks not possessing hierarchical organization or multiple modules. Here we explored how variation in the number of hierarchical levels and modules per level influenced network dynamics and occurrence of LSA. We tested hierarchical configurations of different network sizes, approximating the large-scale networks linking cortical columns in one hemisphere of the rat, cat, or macaque monkey brain. Scaling of the network size affected the number of hierarchical levels and modules in the optimal networks, also depending on whether global edge density or the numbers of connections per node were kept constant. For constant edge density, only few network configurations, possessing an intermediate number of levels and a large number of modules, led to a large range of LSA independent of brain size. For a constant number of node connections, there was a trend for optimal configurations in larger-size networks to possess a larger number of hierarchical levels or more modules. These results may help to explain the trend to greater network complexity apparent in larger brains and may indicate that this complexity is required for maintaining stable levels of neural activation.

## 1. Introduction

Complex systems operate within a critical functional range (Bak et al., 1987), sustaining diverse dynamical states on the basis of their intricate system architecture. Criticality is associated with the phase transition between ordered and chaotic dynamics, and systems tuned to the critical point produce power-law distributions in their dynamics. Recent studies indicate that brain networks also operate close to a critical point. Evidence for this comes, for example, from the observation of neuronal avalanches (i.e., bursts of activity separated by longer periods of relative rest) with a power-law size distribution in cortical slices (Beggs and Plenz 2003), and from time series analysis of EEG data (Freeman et al., 2000) showing that the power spectral density of background activity follows a power law. Critical activity has also been demonstrated in human brain functional networks (Kitzbichler et al., 2009). While its functional significance is still not well understood, it has been suggested that critical dynamics may enhance information processing capabilities of neuronal networks (e.g. Bertschinger and Natschlager, 2004). This idea is supported by work showing that the dynamic range in an excitable network is optimized at criticality (Kinouchi and Copelli, 2006). Given these findings, it is desirable to obtain a better understanding of the conditions for criticality in complex excitable networks.

A necessary precondition for attaining a critical point in intricate neural systems, such as the mammalian cerebral cortex, is that initial network activations result in neuronal activation patterns that neither die out too quickly nor spread across the entire network. Without this feature, activation patterns would not be stable, or would lead to a pathological excitation of the whole brain. What are the essential structural and functional parameters that allow complex neural networks to maintain such a dynamic balance of sustained yet limited activity? Most current models of neural network dynamics focus on maintaining the right balance of network activation and rest through functional interactions among populations of inhibitory and excitatory nodes (Beggs and Plenz, 2003). Alternative balancing mechanisms may be provided by broad external input from neuromodulatory systems or self-sustained neuronal activity (Muresan and Savin, 2007). However, the topology of neural networks may also contribute to critical network dynamics, even in the absence of explicit inhibition. For this reason, we are particularly interested in the relationship of different kinds of neural network topology to the condition of limited sustained activation (LSA). The involvement of inhibitory neuronal populations and other dynamic control mechanisms may then further extend the parameter range for LSA that is provided by principal topological features of the neural network architecture.

Two central topological features of brain networks, in particular of the cerebral cortex, are their modular and hierarchical organization. A modular hierarchical organization of cortical architecture and connections is apparent across many scales, from cellular microcircuits in cortical columns (Mountcastle 1997; Binzegger et al., 2004) at the lowest level, via cortical areas at the mesoscopic scale, to clusters of highly connected brain regions at the global systems level (Breakspear and Stam, 2005; Hilgetag et al., 2000; Kaiser, 2007). The precise organization of these features at each level is still sketchy, and there is exists controversy about the exact organization or existence of modules even at the level of cortical columns (Rakic, 2008; Smith, 2009). Nonetheless, current data and concepts suggest that at each level of neural organization clusters arise, with denser connectivity within than between the modules. This means that neurons within a column, area or cluster of areas are more frequently linked with each other than with neurons in the rest of the network.

The spreading of activity has been modeled for cortical networks (Kötter and Sommer, 2000) and other complex networks with a non-random organization (Pastor-Satorras and Vespignani, 2001). In a previous study of activity spreading through different topologies of excitable networks (Kaiser et al., 2007a), we showed that patterns of limited but sustained activity are well supported by the organization of hierarchical multi-modular networks, but not random or simple small-world networks (Watts and Strogatz, 1998) of the same size. In addition, such properties arose without the need for explicit inhibitory feedback or external input, demonstrating the significant role of network topology in sustaining and limiting neural activation (Latham and Nirenberg, 2004; Roxin et al., 2004).

While our previous study (Kaiser et al., 2007a) demonstrated that a wide range of initial parameters in hierarchical modular networks could result in LSA, it did not clarify whether this range was due mainly to the multi-modular organization of the network or its hierarchical structure. In the present study, we investigated the relation of different hierarchical network configurations to the range of LSA more extensively. The principal type of hierarchical structure, an interconnected set of modules with encapsulated sub-modules without explicit hub nodes, as well as the settings for the dynamic mechanisms were preserved from our previous model. A fixed number of nodes was activated at the beginning of each dynamical simulation. Other nodes became activated when at least $k$ of their directly connected node neighbors were active at the same time. Each active node deactivated in the following time step with probability $v$. Note that this model only assumes initial activation at time step zero, but no ongoing external input or internal random activation.

Our hierarchical topological model reflects general features of brain connectivity at the large and mesoscopic scale, in particular the modularity of neural networks across scales. Nodes in the model are intended to represent cortical columns (Mountcastle, 1997) rather than individual neurons. Connections between columns are modeled as exclusively excitatory, since it is appears to hold that there are no long-distance inhibitory connections within the cerebral cortex (Latham and Nirenberg, 2004). However, nodes can also deactivate (controlled via the model's deactivation probability) due to intrinsic inhibition or exhaustion, as observed for cortical tissue after prolonged firing, for instance in epileptic seizures (Milton and Jung, 2003).

Two main parameters were explored in the hierarchical networks, the number of hierarchical levels and the number of sub-modules at each level (cf. Figure 1). These parameters were varied, while other topological features, such as the probability that any two nodes are connected, or alternatively, the average number of connections per node, were kept constant. We explored whether optimal hierarchical configurations existed, in which the proportion of tested cases with LSA was at a maximum. In addition, we tested whether the parameters for such optimal configurations changed with network size; that is, whether small networks, representing the approximate number of columns as in a small rodent (rat) brain, had different optimal settings than larger cortical networks that might reflect the number of column nodes in larger mammalian (cat) and primate (macaque) brains.

## 2. Materials and Methods
Calculations were performed on a 16-core HP ProLiant server using the Linux version of Matlab R2009a (Mathworks Inc., Natick, MA). Scripts are available at http://www.biological-networks.org.

### 2.1. Anatomical constraints
We investigated if the topology of optimal hierarchical networks, leading to a maximum parameter range of LSA, varied with brain size. For this approach, the number of nodes was set to the number of columns estimated to exist in one cortical hemisphere in different species. The number of columns was estimated from the surface size of one cortical hemisphere in rat (Nieuwenhuys et al., 1998), cat (ibid.), and macaque (Felleman and van Essen, 1991) under the assumption of each (macro-)column occupying 1 mm$^2$. Real columns might be smaller and we elaborate on the role of column size differences across areas and species in the Discussion. We explored three networks with different surface sizes for one hemisphere, rat-like (300 nodes; 3 cm$^2$ surface), cat-like (4,150 nodes; 41.5 cm$^2$ surface), and macaque-like (11,000 nodes; 110cm$^2$ surface). Note that these are very simple estimates based on the assumption that columns in different species are comparable in the basic circuit layout even though the absolute number of neurons may vary (Herculano-Houzel et al., 2008). Either edge density or average number of edges per node (average degree <k>) was kept constant across network sizes. The edge density was set to 1.2%, corresponding to the one chosen in a previous study (Kaiser et al., 2007a) and is close to values of 0.48% for a model of the rat cortex (Kaiser, 2009).

The constraint of a constant average node degree was motivated by comparative studies showing largely constant numbers of connections per neuron across many species (Binzegger et al., 2004; Changizi and Shimojo, 2005; Hellwig, 2000; Schüz and Braitenberg, 2002; Schüz et al., 2005; Striedter, 2004). At the column level, it can also be reasoned that columns are mostly connected with adjacent columns on the cortical surface (short-distance; intra-areal) and only a few columns in different areas (long-distance; inter-areal connections). Due to this presumed homogenous arrangement of cortical networks, the number of connections per column should be independent of the total number of columns in the network. The average degree was set to the arbitrary but fixed number of 50. Note that the actual values for edge density or average degree might differ from the ones chosen here without changing the principal conclusions of this study, as results across different networks were compared qualitatively.

## 2.2. Generating hierarchical networks

Alternative approaches exist for generating a hierarchical network with *m* sub-modules per module and a total number of levels *h*. As a default, we settled on a strategy in which the total number of edges *E* was distributed to the different levels (see Figure 1) with $E_i$ edges on level *i*, so that each level received the same number of edges: $E_i = E / (h+1)$. This model, which was used throughout the study, preserved a constant number of edges when the number of levels or sub-modules within modules was varied.

The network was generated beginning with the highest level and adding modules to the next lower level with random connectivity within modules. The resulting networks were similar to the ones produced by an alternative procedure (Sporns, 2006), but differed in the generating algorithm (we present pseudocode here, the actual Matlab algorithm is available online at http://www.biological-networks.org):

```
for i from 0 to h-1                                    for all hierarchical levels
    A_i := (m-1)/m^(i+1)                               proportion of the adjacency matrix occupied by
                                                       one module at the current hierarchical level
    p_i := E_i / (N^2 A_i)                             edge density within a module
    N_i := ⌊ N * (1/m)^i ⌋                             number of nodes in a module
    N_c := N / N_i                                     number of modules at the current level

    for j from 1 to N_c
        c_i := random graph with N_i and edge density p_i    random N_i × N_i graph with edge density p_i
        r_0 := 1 + (j-1) N_i                                  first node of the module
        r_1 := r_0 + N_i – 1                                  last node of the module
        CIJ(r_0 to r_1, r_0 to r_1) := c_i                    part of the matrix CIJ is replaced with c_i

for i from 1 to N
    CIJ(i, i) = 0                                      remove connections across the diagonal (loops)
```

The algorithm involves the following steps: Starting with an empty adjacency matrix CIJ, modules at hierarchical level *i* are added starting with modules at the hierarchical level h=0 (global network). The ratio of matrix elements that represent potential connections within any module $A_i$ depends on the hierarchical level *i* and the number of sub-modules per module *m*. Based on $A_i$ and the number of edges at that hierarchical level, $E_i$, we can determine the probability $p_i$ that any two nodes within a module are connected (edge density within modules). Next, the number of nodes in each module $N_i$ is calculated leading to the total number of modules $N_c$ at that level. Then, each module given as a random graph with edge density $p_i$ is inserted in the adjacency matrix *CIJ*. Finally, edges along the diagonal (loops) are removed from the network. Due to this removal, the total number of edges might be slightly lower than the desired number of edges. In these cases, additional edges are added randomly to the network to generate the desired total number of edges and edge density (not shown in the pseudocode; however, see Matlab routine online).

We explored hierarchical networks with different numbers of hierarchical levels, *h* (scales), and numbers of sub-modules at each level, *m* (granularity). A network without hierarchical levels forms a random network, with one level a 'flat' modular network, two levels a network with modules and sub-modules, and so on (Figure 1).

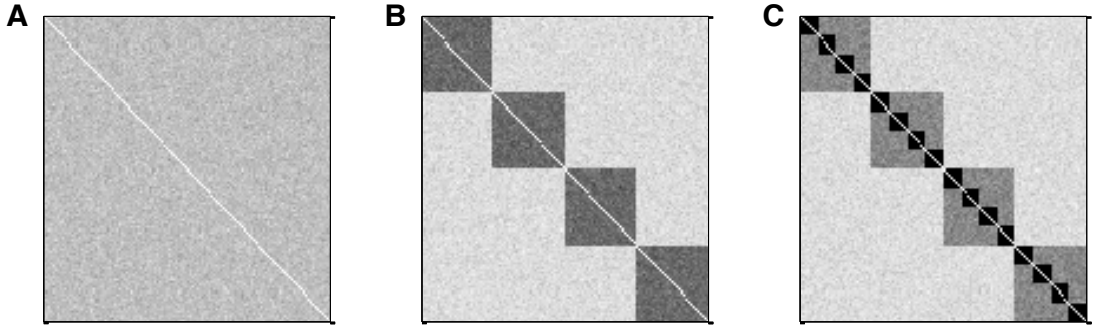

**Figure 1: Overview of variation of granularity and scales in the explored hierarchical modular networks.** The plots show the outcome of one hundred realizations of networks with 128 nodes and 4096 directed edges. Gray level shading of the adjacency matrix indicates the frequency with which an edge was established (white: never established; black: established in all 100 generated networks). (**A**) Random networks without hierarchical structure, resulting from $h = 0$ (number of hierarchical levels) and $m = 0$ (number of sub-modules); (**B**) Flat modular networks with four modules, resulting from $h = 1$, $m = 4$; (**C**) Hierarchical modular networks with $h = 2$, $m = 4$. Note that each hierarchical level contains the same number of edges, resulting in 16 modules at the lowest hierarchical level in (C), which possess the highest edge density.

### 2.3. Spreading model
A basic spreading model (Newman, 2005) was modified to simulate the propagation of activity through the network. This dynamic model was identical to the one used in (Kaiser et al., 2007a).

The simulation operated in discrete time steps, with nodes being in one of two states, active or inactive. We used a simple threshold model for activity spreading where a number *i* of randomly selected nodes was activated in the first step. At each subsequent time step, inactive nodes became activated if at least *k* neighbors were currently active (neighbors of a node are nodes to which direct connections exist). Activated nodes could become inactive with probability *v* in the next time step, or otherwise stayed active.

An additional parameter was the extent of localization of the initial activation, $i_0$. For initialization, *i* ($i \leq i_0$) nodes among the nodes 1 to $i_0$ were selected randomly and activated in the first time step. The networks nodes were numbered consecutively. For instance, for a network where the largest modules at the highest level contained 100 nodes and where each module contained 10 sub-modules with 10 nodes each, by setting $i_0$ to 10, 20 or 100, the first sub-module, the first two sub-modules, or the first module, respectively, were activated during initialization. Thus, *i* determined the number of initially active nodes while $i_0$ controlled the localization of initial activations, with smaller values resulting in more localized initial activity.

## 2.4. Calculating the average range of limited sustained activity

We systematically explored the network activation resulting from different settings of the initial node activation and localization parameters. Persistent contained activity in hierarchical networks (e.g., intermediate-level trace in Fig, 2A) existed for a wide range of initial localization and activation parameters (indicated by gray filled circles in Figure 2B).

We also explored if the results were robust for variations in the dynamic model parameters $k$ and $v$, by using a Monte Carlo approach in which, for each pair of $k$ and $v$, spreading simulations with randomly chosen parameters $i$ (number of initially activated nodes) and $i_0$ (localization) were tested (Fig. 2B). A trial was considered to show sustained activity if at least one but at most 50% of the nodes were activated at the end of the simulation (after 200 steps). In our experience, activity did not further die out or spread through the whole network if such an activity level was reached at the end of the simulation. For each pair of spreading parameters $k$ and $v$, the average proportion of cases for which sustained activity occurred was charted (Figure 2C). This proportion is specified by the ratio of gray filled circles relative to all data points in Figure 2B. The threshold $k$ ranged from 1 to 9 (step size 2), while the deactivation probability $v$ ranged from 10% to 90% (step size 20%). Therefore, the average ratio over all entries in Figure 2C reflected the size of the parameter space for a given network topology which could give rise to LSA, taking into account the initialization parameters $i$ and $i_0$ as well as the dynamic model parameters $v$ and $k$.

We tested the proportion of cases with sustained activity for different hierarchical configurations. These configurations varied in the number of hierarchical levels, from zero for random networks to four, and the number of sub-modules into which each module was divided for creating the next-lower hierarchical level. For each configuration, different values for the threshold $k$ and the deactivation probability $v$ were tested, in that for each ($k$, $v$) pair, 200 runs were performed and the network state was observed after 200 time steps leading to a classification as dying-out, sustained, or spreading activity. For these 200 runs, the number of initially activated nodes $i$ and the localization parameter $i_0$ was chosen randomly (see range in Figure 2B). The average proportion of sustained activity cases for each configuration was plotted as gray-scale value for Figure 4 and the subsequent figures.

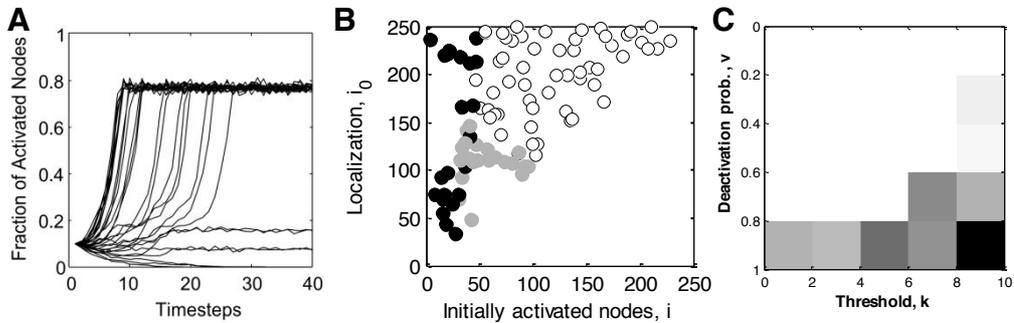

**Figure 2: Determining the parameter range of limited sustained activity (schematic overview).** (**A**) For several trials (shown here: 30 runs), it was tested whether activity spread through the whole network (here: activating 80% of all nodes), died out (all nodes becoming inactive), or was sustained at an intermediate level (here: activating 10% or 20% of all nodes). Note that even during complete spreading, not all the nodes were constantly active, due to the inactivation probability $v$ specified in the dynamic model. (**B**) Simulations were run for different combinations of the number of initially activated nodes $i$ and the localization parameter $i_0$. For each run, the simulated activity died out (●), spread through the whole network (o) or was sustained within a limited compartment of the network (●). (**C**) The parameter space of simulations was further explored for different combinations of deactivation probability $v$ and activation threshold $k$. Gray levels for each parameter combination in the diagram reflect the percentage of cases giving rise to LSA (from subplot B). The average value across all entries was taken as the final measure of the parameter range of functional criticality for a particular network topology. It reflects the average proportion of limited sustained activation cases obtained across all parameter settings for a given hierarchical modular network.

## 2.5. Inaccessible parameter range

For all network sizes, a variety of hierarchical configurations could not be realized, due to the limited network size (regions indicated by horizontal lines in Figure 3 and subsequent figures). Inadmissible configurations were those where the smallest module at the bottom level would have contained more edges than there were edges possible between module members; that means where $N_c$ ($N_c$-1) < $E_c$ ($N_c$: number of nodes; $E_c$: number of edges in the smallest module). Note that the number of sub-modules per module was varied in steps of two (2, 4, 6, 8,…). Variations by a different step size might have produced more clearly apparent differences in inaccessible configurations between small (300 nodes) and large (11,000 nodes) networks.

## 3. Results
### 3.1. Expiring, limited sustained and completely spreading activity patterns

How does activity change over time for different parameter settings? In Figure 3 we give examples for different outcomes in a network with 512 nodes, two hierarchical levels, and eight modules with eight sub-modules per module. Modules are represented by gray shading where the individual gray levels represent sub-modules. Blue dots indicate that a node is active at a certain time step.

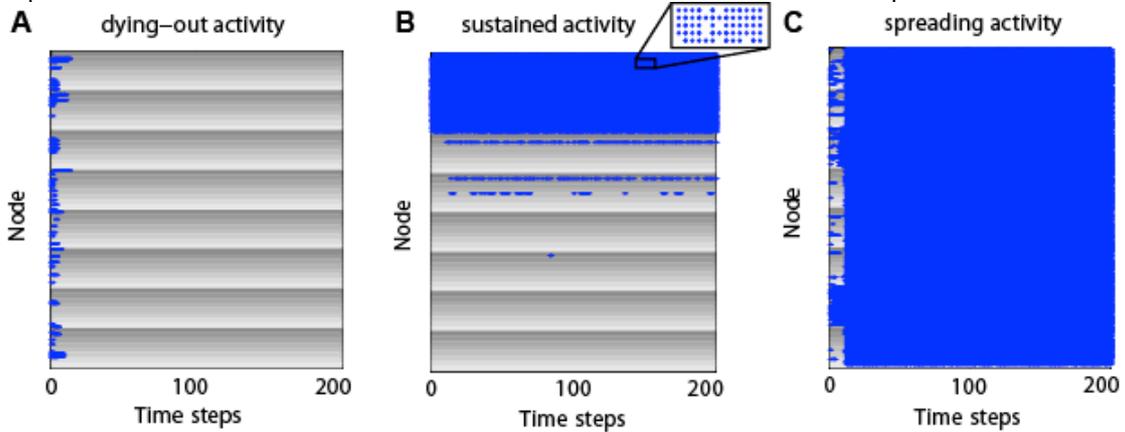

**Figure 3: Examples for neural dynamics for different simulation outcomes.** Gray shading represents modules and individual gray levels represent different sub-modules. Nodes which are active at a time step are represented as blue dots. (**A**) Expiring (dying-out) activity. (**B**) Limited sustained activity. Although some modules appear completely activated, nodes can be inactive at various time steps due to the inactivation probability (inset). (**C**) Completely spreading activity.

For expiring activity (Figure 3A), initial activity quickly died out as active nodes became de-activated and not enough active neighbors existed to sustain the activity. For limited sustained activity (Figure 3B), modules and sub-modules became activated indicating that a critical number of neighbors of a node were active and able to (re-) activate a node. For completely spreading activity (Figure 3C), activity that was initially contained in one module or several sub-modules managed to spread to other parts of the network and quickly led to complete network activation. This time-course of an early focus of activity with a rapid spread to the whole network may be compared to the generalizations of seizures in epilepsy patients. Note that the blue lines in Fig. 3A as well as the large blue areas in Fig. 3B and C also contain nodes which are not active; however, these nodes are not visible in the figure due to the dot size and image resolution.

### 3.2. Topological and small-world properties of hierarchical networks

For all tested network sizes, the generated hierarchical networks ($h \geq 1$) possessed characteristics of small-world networks (Watts and Strogatz, 1998), in that the clustering coefficients (the average frequency with which neighbors of a node are directly connected) were much higher than for same-size Erdös-Rényi random networks (Erdös and Rényi, 1960), whereas the characteristic path lengths (the average number of connections on the shortest path between any two nodes) remained comparable to those for random networks of the same size (Tables 1 and 2). Note that networks with only one hierarchical level represent the special case of simple modular networks.

**Table 1. Graph and small-world characteristics of hierarchical networks with constant edge density.** The number of edges $E$ for a given number of nodes $N$ was chosen such that the edge density remained 1.2%. Networks with two hierarchical levels and four sub-modules per module are shown (cf. Figure 1C). $C$ and $C_{rand}$: clustering coefficients of the hierarchical and random network. $L$ and $L_{rand}$: Characteristic path lengths of the hierarchical and random network. $SW$: small-world index $(C/C_{rand}) / (L/L_{rand})$.

| $N$ | $E$ | $C$ | $C_{rand}$ | $L$ | $L_{rand}$ | $SW$ |
|---|---|---|---|---|---|---|
| 300 | 1,080 | 0.025 | 0.012 | 23.6 | 25.0 | 2.21 |
| 512 | 3,146 | 0.024 | 0.012 | 5.1 | 4.6 | 1.80 |
| 4,150 | 206,670 | 0.023 | 0.012 | 2.6 | 2.5 | 1.84 |
| 11,000 | 1,452,000 | 0.023 | 0.012 | 2.3 | 2.2 | 1.83 |

The characteristic path length for the case of constant edge density (Table 1) was particularly high for the 300-node network. This is due to the low edge density of 1.2%; small networks with low edge density exhibit fewer alternative pathways than larger random networks with the same edge density. Therefore, the path length decreases when more edges are added, as for the larger networks. This behaviour resembles the behaviour of random networks where the characteristic path lengh $L \sim \ln N / \ln <k>$, where $N$ is the number of nodes and $<k>$ is the average node degree (Albert and Barabási, 2002; Costa et al., 2007). All networks, however, show features of small-world networks (Watts and Strogatz, 1998). The clustering coefficient for random networks, $C_{rand}$, was the same for all network sizes. For random networks, the clustering coefficient is the same as the edge density; that means, the probability that neighbors of a node are connected is the same as the probability that any two nodes are connected. As the edge density is kept constant for all network sizes, $C_{rand}$ remains constant at that value as well. The extent of a small-world organization can be characterized by the small-world coefficient $SW = (C/C_{rand}) / (L/L_{rand})$ (Humphries et al., 2006; Humphries and Gurney, 2008). The index $SW$ is around 2 indicating a small-world organization of these networks. Whereas SW is 2.2 for a small network size of 300 nodes, it remains at a lower level of 1.8 for larger networks.

**Table 2. Graph and small-world characteristics of hierarchical networks with constant node degree.** The number of edges $E$ for a given number of nodes $N$ was chosen such that the average number of connections (node degree) was 50. $C$ and $C_{rand}$: clustering coefficients of the hierarchical and random network. $L$ and $L_{rand}$: Characteristic path lengths of the hierarchical and random network. $SW$: small-world index $(C/C_{rand}) / (L/L_{rand})$.

| $N$ | $E$ | $C$ | $C_{rand}$ | $L$ | $L_{rand}$ | $SW$ |
|---|---|---|---|---|---|---|
| 300 | 15,000 | 0.227 | 0.167 | 1.8 | 1.8 | 1.36 |
| 512 | 25,600 | 0.163 | 0.098 | 1.9 | 1.9 | 1.66 |
| 4,150 | 207,500 | 0.023 | 0.012 | 2.6 | 2.5 | 1.84 |
| 11,000 | 550,000 | 0.009 | 0.005 | 2.8 | 2.8 | 1.80 |

For constant average node degree $<k>$ (Table 2), the average path length increases with network size. Smaller networks with 300 and 512 nodes show a considerably lower path length compared to constant edge density. Again, all networks displayed features of small-world networks (Watts and Strogatz, 1998). The small-world index $SW$, however, was lower for small network sizes of 300 and 512 nodes compared with the scenario of constant edge density.

### 3.3. Optimal hierarchical configurations for LSA in a small network
#### 3.3.1. Variation of sustained activity and topological measures
As a first test, we explored the link between hierarchical organization and the parameter range of LSA for different configurations of a network with 512 nodes and, on average, 50 connections per node (Figure 4A). The parameter range for LSA tended to increase with the number of sub-modules at each hierarchical level (along rows in Fig. 4A). The maximum range of LSA occurred for one hierarchical level and the largest possible number of sub-modules per module.

In this and all following plots (Fig. 5-6), regions with horizontal lines indicate hierarchical configurations that cannot be realized, as some modules would need to contain more edges than can be fitted between members of that module. These cases were detected whenever $N_c$ $(N_c-1) < E_c$ ($N_c$: number of nodes; $E_c$: number of edges in the smallest module); that means the number $E_c$ of edges that needed to be established was higher than the number of possible edges in a module, $N_c$ $(N_c-1)$. Note also that the gray levels indicating proportion of cases were normalized so that white regions represent the minimum and black regions the maximum value for each plot.

Due to the network generation algorithm, modules at the lowest level of the hierarchy had the largest edge density (cf. Figure 1). We used this effect to test if limited sustained activation patterns were facilitated by more densely connected bottom modules in the network (Figure 3B). Interestingly, there existed no clear relation of the density with sustained activation: whereas both maximum edge density and sustained activity probability increased with the number of sub-modules for a network with two hierarchical levels (Figure 4A, B), the relation was less clear for larger numbers of hierarchical levels.

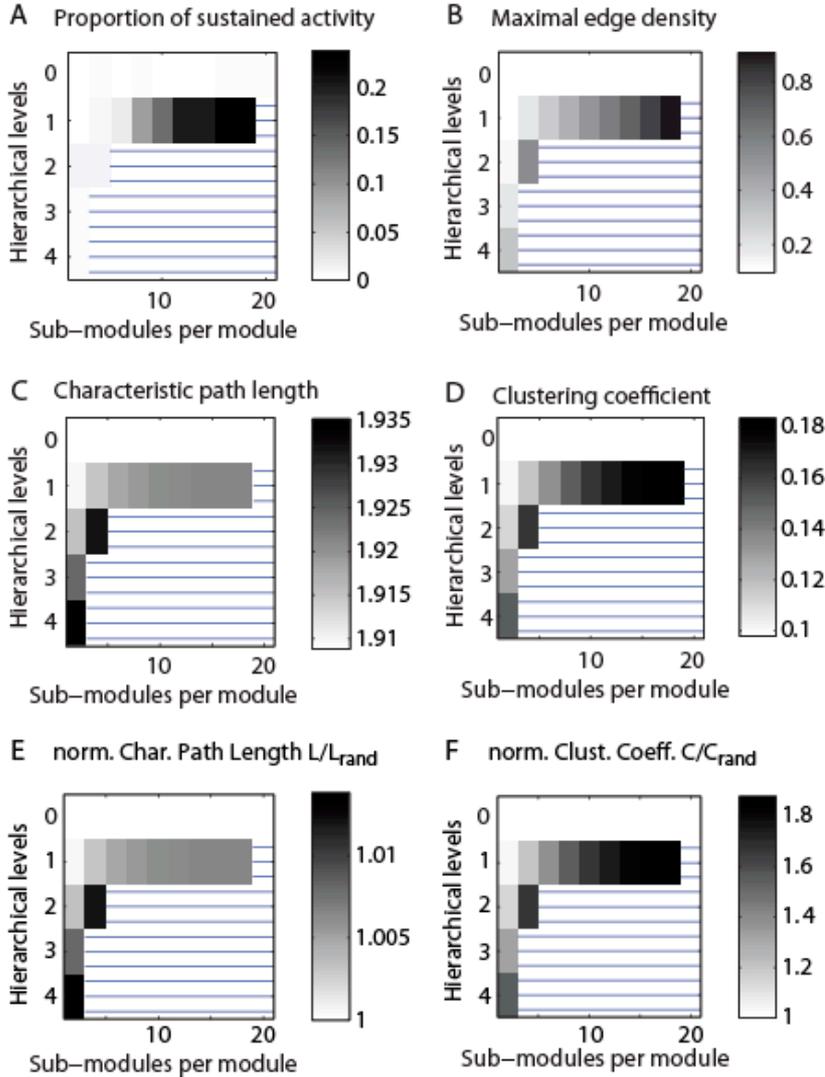

**Figure 4: Range of limited sustained activity for different hierarchical configurations of a small network.** Shown is the parameter range of limited sustained activation and of topological features for a network with 512 nodes and average node degree of 50. Regions blocked by horizontal lines indicate configurations that were not admissible (see Methods). Parameters were explored for 1,000 runs of each set of spreading parameters $k$ and $v$, while the number of initially activated nodes $i$ and the localization parameter $i_0$ varied for each run. (**A**) Average of the number of parameter settings leading to LSA. (**B**) Maximum edge density based on the most highly connected modules (modules at the lowest level of the respective hierarchy). (**C**) Characteristic path length of the networks. (**D**) Average clustering coefficient of the networks. (**E**) Normalized characteristic path length (divided by the value for Erdös-Rényi random networks). (**F**) Normalized average clustering coefficient (divided by the value for Erdös-Rényi random networks).

How are small-world properties linked to the different hierarchical configurations? The characteristic path length (Figure 4C) appeared to show lower values when two or more hierarchical levels existed in the network, but the values were in a narrow range of 5.3 to 6.4. The clustering coefficient (Figure 4D) increased with the number of levels and the number of sub-modules per module. The characteristic path lengths of the hierarchical networks were comparable to those of Erdös-Rényi random networks (Figure 4E) whereas the clustering coefficient was higher than in random networks (Figure 4F). As the normalized path length is around 1, the SW index has a similar value as the normalized clustering coefficient. Given large SW indices, the networks possessed features of small-world networks (Watts and Strogatz, 1998).

### 3.3.2. Control calculations
We tested several parameters that were used for generating hierarchical networks. Networks consisted of 512 nodes and, on average, 50 connections per node. Appendix A contains a full description of these control calculations including additional figures.

*Varying the number of edges for different hierarchical levels*
By default, the number of edges for each hierarchical level was set to be equal, that means, $E_i$ was the same for each hierarchical level $i$. Here, we tested sustained activity patterns for varying numbers of edges per level. We observed two cases: (a) a decrease of the number of edges with each hierarchical level or (b) an increase of the number of edges with each hierarchical level. The absolute level of sustained activity was lower in case (a) and higher in case (b) compared to the original setting (cf. Figure A1).

*Varying the parcellation for different hierarchical levels*
By default, we used the same parcellation of modules at each level; that means if a module consisted of two sub-modules for the highest level, this condition would be the same for all other levels of the network hierarchy as well. Here, we also tested varying the parcellation into sub-modules depending on the hierarchical level. Again, we tested two cases: (a) creating fewer sub-modules for higher hierarchical levels ("decreasing parcellation") or (b) creating more sub-modules for higher hierarchical levels ("increasing parcellation"). Configurations with a high proportion of LSA cases for decreased as well as increased numbers of sub-modules for each hierarchical level remained comparable with the original calculation. However, for the "increasing parcellation" type, the overall proportion of LSA cases increased, extending the maximum probability of sustained activity from 0.23 to 0.42 (cf. Figure A2).

*Number of cases close to 50% activation threshold for classification as sustained*
In additional simulations, we tested how close the final activity was to the threshold used for classification as a case of LSA. Indeed, final activity levels close to the 50% threshold could occur. However, final activity levels were around 10-20% for most configurations producing a high number of LSA cases. This indicates that configurations leading to a high proportion of LSA cases were not substantially affected by the threshold (cf. Figure A3).

*Topologies leading to expiring, limited sustained, and completely spreading activity*
As a default, we investigated the distribution of LSA cases depending on the hierarchical network organization. In additional simulations, we also observed the distribution of the other two possible simulation outcomes: activity dying out before the end of the simulation and final activation of more than 50% of the nodes, which was classified as complete spreading. For a random network (zero hierarchical levels), both dying-out and complete spreading occurred in 50% of the cases. However, when more than one hierarchical level was present, complete spreading activity occurred more often than dying-out. For one hierarchical level, outcomes depended strongly on the number of sub-modules per module. The cases with expiring activity formed 50% of the cases for two sub-modules, but decreased with the number of sub-modules (cf. Figure A4).

*Varying the edge density*
The pattern of sustained activity remained comparable to the default settings even when the edge density differed from the original value of 10% for a network with 512 nodes and an average node degree of 50. The maximum proportion of cases with LSA varied between 0.172 for decreased edge density (5%) to 0.238 for increased edge density (20%). The relative distribution of case for networks with one hierarchical level was similar across edge densities (cf. Figure A5).

## 3.4. Scaling of optimal hierarchical configurations for LSA with network size

Two different scaling scenarios were explored. In the first one, the global edge density of the networks was kept constant (at 1.2%) while the average number of connections per node varied; in the second scenario, the average node degree was kept constant (at 50 connections per node) while the networks' edge density varied.

### 3.4.1. Constant edge density

In the first approach, the probability that any two nodes (representing cortical columns) in the network were connected was, on average, 1.2%. Connection density was larger within modules and lower between modules; however the global average remained constant, independent of the hierarchical configuration.

Given a constant setting for testing neural activation across network sizes, the 'rat-size' network showed sustained activity for a wide variety of hierarchical configurations (Figure 5A). Surprisingly, for the larger 'cat-size' network, a smaller variety of hierarchies existed that could generate LSA (Figure 5B). However, for these configurations, sustained activity occurred in up to 25% of the explored parameter settings, whereas it occurred only in up to 3% of the tested settings for the rat-like network (cf. scale of activation range). For the even larger 'macaque-size' network, the maximum range of limited sustained activation (up to 4% of tested parameter settings) was as low as for the 'rat-size' network (Figure 5C), while overall, the variety of hierarchical configurations that resulted in limited sustained activity was also lowest for the macaque-like network.

These results indicate that the number of possible hierarchical configurations (resulting from combinations of the number of levels and number of sub-modules) leading to LSA decreased with increasing network size. Only a few hierarchical configurations appeared suitable for producing LSA in all network sizes. Such configurations typically combined an intermediate number of levels with a large number of sub-modules (Figures 5B, 4C). Interestingly, the combination of a large number of hierarchical levels with a small number of sub-modules proved ineffective for supporting LSA in the larger-size network (Figure 5C). For large networks the best strategy for achieving sustained activity was provided by an arrangement of two hierarchical levels containing the largest possible number of modules and sub-modules.

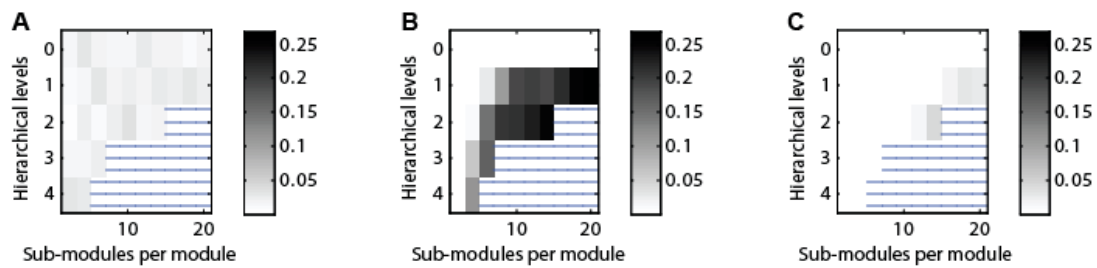

**Figure 5: Scaling of optimal configurations with network size for constant global edge density.** Proportion of cases showing LSA (averaged over 200 generated networks for each configuration) in (**A**) 'rat-size' networks with 300 columns, (**B**) 'cat-size' networks with 4,150 columns, (**C**) 'macaque-size' networks with 11,000 columns.

### 3.4.2. Constant average node degree

The results obtained under the constraint of a constant edge density (Section 3.4.1) suggested that configurations for LSA were harder to attain in large as well as small networks. Only networks of an intermediate size appeared to result in a large variety of hierarchical networks possessing a wide parameter range for LSA. In an alternative approach, we also tested optimal configurations of networks of different sizes under the constraint that the average number of connections per column, rather than the probability that any two columns are connected (edge density), was kept constant. For this approach, the number of edges was set to 50 times the number of nodes, leading to an average node degree of 50 in all networks, albeit with variation for individual nodes.

Under these conditions, LSA in the 'rat-size' networks arose mostly in networks with one hierarchical level, and for an increasing number of sub-modules (Figure 6A). Both the 'cat-size' and the 'macaque-size' networks possessed a similar, large range of hierarchical configurations showing sustained activity (Figure 6B, C). All networks demonstrated that cases of LSA increased with the number of sub-modules per module. Whereas the range of hierarchical configurations differed between the 'rat-size' and the 'cat-size' or 'macaque-size' networks, the maximum range of sustained activity was comparable for all sizes with 15-30%. A constant number of connections per node, therefore, permitted a wide range of optimal hierarchical configurations for LSA even if the network size increased.

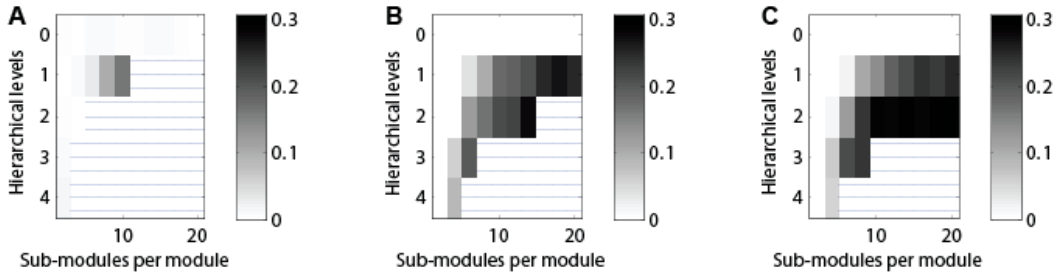

**Figure 6: Scaling of optimal configurations with network size for constant average node degree (<k> = 50).** Proportion of cases showing LSA (averaged over 200 networks generated for each configuration) in (**A**) 'rat-size' networks with 300 columns, (**B**) 'cat-size' networks with 4,150 columns, and (**C**) 'macaque-size' networks with 11,000 columns.

## 4. Discussion

This study investigated an essential precondition of criticality in neural systems, the capability of neural networks to produce limited sustained activation patterns following an initial activation. We addressed this question by simulating the spreading of neural activity and systematically varying model parameters and network topology in hierarchical modular networks, which are inspired by the organization of biological neural networks across scales. Our previous study (Kaiser et al., 2007a) demonstrated that hierarchical cluster networks possess a large parameter range leading to LSA, in contrast to random and non-hierarchical small-world networks. Here we expanded this analysis by varying the number of levels and sub-modules in hierarchical networks and scaling their size within two alternative scenarios, constant edge density or constant average node degree. This study demonstrated, first, that LSA patterns are supported by a variety of parameter settings for hierarchical modular networks, combining different numbers of hierarchical levels with varying numbers of sub-modules per level; second, that for the same network size and the same number of sub-modules, networks with a larger number of levels resulted in a wider range of LSA, while for the same number of hierarchical levels a larger activity parameter range was produced by increasing the number of sub-modules; and third, that a high level of sustained activity was attainable across network sizes for a constant average node degree, but not for constant edge density.

The present results provide a proof of concept for three points. First, hierarchical network configurations lead to different levels of sustained activity independent of global topological properties, such as characteristic path length or clustering coefficient. Therefore, the identification of an optimal network configuration associated with a maximum level of LSA is a suitable target for evolutionary graph optimization. Second, only specific hierarchical network arrangements can be realized for a limited network size. Even for the human brain with an estimated number of 125,000 columns per hemisphere under the assumptions made in Section 2.1, only a small fraction of potential hierarchies can be realized within the current framework. For the 'human-size' network, three hierarchical levels with up to 18 sub-modules and four levels with up to 8 sub-modules are possible. These limits are beyond the ones of the 'macaque-size' network which maximally allowed six and four sub-modules for three and four hierarchical levels, respectively. Therefore, simple combinatorics suggest that it is easier to vary the number of modules at each level than to increase the number of levels for larger brains.

Third, the number of configurations which lead to sustained activity decreases with network size if the edge density remains constant, but remains large even for large network sizes if the average number of connections per node is kept constant. This model finding corresponds to the observation from comparative studies that the number of connections of a neural node (e.g. the number of synapses of a neuron) rather than the ratio of connections (e.g. being connected to 10% of all neighbors) largely remains constant across species with different brain sizes (Ringo, 1991; Schüz and Demianenko, 1995; Striedter, 2004; Zhang and Sejnowski, 2000). Moreover, for constant average node degree, the optimal configurations in larger networks tended to possess more hierarchical levels, suggesting a beneficial contribution of the more intricately structured topology in larger neural networks to dynamical stability. There are indications from compilations of biological neural connectivity (e.g., www.cocomac.org) that support this model finding. For example, cluster analyses suggest that primate cortical connectivity is structured on more levels than connectivity in smaller brains, if one considers that there exist primate visual 'streams' (Young 1992; Hilgetag et al. 2000), that is, subdivisions of the visual network module that are apparently absent in the rat (Burns et al., 2000) or cat network (Scannell et al. 1999; Zamora-Lopez et al., this issue.) Moreover, there are generally more modules in larger brains, if cortical areas can be considered as modules.

The finding of increased hierarchical structure in larger networks may appear counterintuitive given that there are limits on the number of hierarchical levels even in large networks, as discussed above. However, an appropriately large number of levels may be a necessary constraint for sustaining activity. If the number of modules in a large network was increased without increasing the number of levels, then, in principle, it would be easy to activate each module. However, activation of the global network may be prevented by dispersion of the activity across the entire network, which means that there may not exist enough projections into each of the individual modules to activate them. Similarly, if there are few large modules, activity may be dispersed *within* the modules. In order to establish a balance between the number and size of modules in large networks, additional levels need to be created, as confirmed by the modeling results.

The hierarchical network topology we explored reflects the distributed multi-level modularity that is considered a central feature of biological neural networks. Neural networks show strong modularity across many levels of scale, ranging from cellular neuronal circuits and neural populations organized in cortical columns (Mountcastle, 1997) to communities of closely linked areas at the systems level (Breakspear and Stam, 2005; Hilgetag et al., 2000). Smaller modules are nested within larger ones, such as columns within an area, which itself is a module in a large-scale brain division, such as the visual system. Another important feature of complex networks that has been discussed widely is the existence of hub, that is, nodes possessing a significantly larger number of links than the majority of nodes in a network (Albert & Barabasi 2002, Ravasz et al., 2002). However, it is difficult to identify nodes in the brain that integrate modules across scales (with the potential exception of unspecific neuromodulatory systems, such as the serotonergic system) and act as global hubs. While there are hub-like nodes in neural networks (Kaiser et al., 2007b; Sporns et al., 2007), they may not act globally, such that most projections in the network originate from, or converge on, a central node. This topology is different from 'centralistic' networks where most nodes are linked to hubs (Ravasz et al., 2002) and which may be more suitable for representing large-scale biochemical networks. However, the detailed investigation of biological neural topologies needs to be continued, since modeling studies have shown a strong impact of topology on network dynamics. For instance, networks which contain hubs may support different modes of activity propagation than hub-less modular networks (Hütt and Lesne, 2009; Muller-Linow et al., 2008).

The present study was set up under several simplifying assumptions, in order to provide general insights into the relationship between hierarchical neural topology and activation patterns. This approach resulted in a number of model limitations. First, nodes representing columns were assumed to be uniform building blocks, whereas actual column organization (layer structure and number of neurons) in the brain might differ across regions (Hutsler et al., 2005) as well as species (Herculano-Houzel et al., 2008), potentially leading to additional constraints on hierarchical network organization. Moreover, the internal organization of columns (self-loops) was not explicitly part of the modeled networks. However, it was represented through the node activation rule: an active node could remain active for the following time step given a sufficient number of active neighbors and potential collaterals going back to the neuron itself. For each time step, the deactivation probability determined whether an active node became inactive. The strength of self-loops was therefore implicitly represented in this deactivation probability with lower likelihood of deactivation for more frequent self-loops. Second, specifically organized populations of inhibitory neurons within columns might additionally influence global network dynamics. Thus, future models could incorporate more detailed biologically realistic mechanisms for reducing activity at the neuronal level, instead of the presently employed phenomenological deactivation probability. Third, the parcellation of modules into sub-modules for each level was treated as symmetric, that means, when a module is split into sub-modules, each module has the same size. It will be important to test asymmetric parcellation of a module into smaller and larger sub-modules in future studies. Finally, the model considered neural network behavior in the absence of external inputs, except for the initial activation. Therefore, the current findings may particularly apply to situations where there is limited external input to the brain, such as during sleep or early development. The results also relate to the organization of neural dynamics associated with the 'default mode' or 'resting state' of the brain (eg, Raichle and Snyder, 2007). The role of external inputs should be addressed in future studies, which could also investigate if there is a difference in the processing of external stimuli by networks that are optimal for LSA and those that are not. In addition, the edge density at the lowest level (Figure 4B) could in some cases be higher than 50% which is unlikely in biological neuronal networks.

In this study, we varied the network size to represent networks of columns of a hemisphere in a 'rat-size' (300 nodes), 'cat-size' (4,150 nodes), and 'macaque-size' brain (11,000 nodes). For large networks and constant edge density, two hierarchical levels with the maximum possible number of sub-modules per module appeared to provide the best strategy for achieving sustained activity (cf. Fig. 5). These multi-level configurations often resulted in a high density of connectivity within modules at the lowest level. Such high edge densities are theoretically possible, but only realized to some extent in biological neural systems. At the global level of human fiber tract connectivity between brain regions, for example, edge densities around 46% can be reached (Honey et al., 2007). Within columns, the connection frequency between any two neurons is around 16% (Douglas and Martin, 2007) but around 35% for neurons from the same cell lineage (Yu et al., 2009). Given constant edge density, the range of feasible hierarchical configurations — that is, the degrees of freedom for evolving neural network architectures —appeared to decrease with larger network sizes. This was due to the fact that the number of neighbors of a node increased with network size. Since the probability for connecting a node to other nodes (the edge density) remained constant, nodes were connected to a larger number of nodes when the number of potential neighbors increased with network size. The larger number of connected neighbors in larger networks also meant that the (absolute) threshold for activating a node was more easily reached. In such cases, activity was harder to contain and more likely spread through the whole network.

Using the constraint of a constant average degree, on the other hand, enabled a wider range of hierarchical configurations with up to 30% sustained activity cases across network sizes (cf. Fig. 6). Such scalability with network size might be beneficial both for ontogenetic and phylogenetic development. Using a constant number of connections per node, rather than a constant edge ratio, across species appears to have several benefits (Changizi, 2001). First, reducing edge density is necessary due to the limited volume available for white matter fiber tracts (Ebbesson, 1980; Karbowski, 2001; Striedter and Northcutt, 2006). For constant edge density, a brain with two times as many columns would contain four times as many connections, quickly increasing brain volume. Second, the setting of a constant number of connections per node provided a setup for sustained activity in different brain network sizes. This might mean that sustained activity can occur for different brain sizes during evolution, if they are appropriately hierarchically structured. Third, hierarchical structuring may also provide the functional stability of LSA in the developing brain. At early stages of ontogenetic development, neural networks generally have few modules and few nodes. During development, more modules, nodes, and hierarchical levels are established (Robinson et al., 2009). Therefore, sustained activity can occur continuously through different stages of development and brain network growth. However, the hierarchical organization is not the only mechanism that can sustain activity during development; early neuronal mechanisms include, for example, spontaneous activity such as retinal waves (Hennig et al., 2009; Sernagor et al., 2006) or the early excitatory function of GABA (gamma aminobutyric acid).

Finally, whereas several earlier studies have explored spatial (e.g. brain volume) or topological (e.g. characteristic path length, Kaiser and Hilgetag, 2006) constraints on brain organization, the present study focused on dynamic constraints, specifically the necessity of brain dynamics to subsist at a sustained yet limited level of activity. Alternative or additional dynamic constraints that may be relevant for this phenomenon could be synchronous activity (Konig et al., 1995; Masuda and Aihara, 2004; von der Malsburg, 1995), or functional attributes such as multi-modal integration, functional complexity (Sporns et al., 2000), information propagation, or processing speed. An accessible parameter range for sustained limited activity is a necessary condition for criticality, but does not in itself guarantee it. Criticality has been interpreted as an abolishing of length scales, that is, the coexistence of dynamical processes at all scales. We saw examples for this phenomenon in activation patterns at LSA where modules and sub-modules of different sizes were activated together. Non-LSA conditions, by contrast, produced only the trivial states of activating all or none of the network nodes. It will be particularly interesting to see how networks optimized with respect to functional diversity are related to networks having optimal range for LSA. A possible link between these two properties was suggested by an earlier analysis showing that the number of significantly repeating activation patterns is maximized at the critical point (Haldemann and Beggs, 2005).

## Disclosure/Conflict of Interest
The authors declare that the research was conducted in the absence of any commercial or financial relationships that could be construed as a potential conflict of interest.

## Acknowledgments
M.K. was supported by WCU program through the National Research Foundation of Korea funded by the Ministry of Education, Science and Technology (R32-10142), the Royal Society (RG/2006/R2) and the CARMEN e-science project (www.carmen.org.uk) funded by EPSRC (EP/E002331/1).

## Appendix A: Control experiments
We tested several additional simulation parameters in the following sections. As in section 3.3, networks consisted of 512 nodes and 50 connections per node.

**Varying the number of edges for different hierarchical levels**
In the default settings, the number of edges $E_i$ was set to be the same for each hierarchical level $i$. Additionally, we tested sustained activity patterns where the number varied. We considered two cases: (a) a decrease of the number of edges with each hierarchical level or (b) an increase of the number of edges with each hierarchical level. The change followed a function where the number of $E_i$ at level $i$ was given by $E_i = s^i E_c / C$, where $s$ is a scaling factor of 2/3 for decreased and 3/2 for increased number of edges per level. The parameter $E_c = E/L$ is the number of edges in a network with $E$ edges and $L$ levels, which was used for the original calculation and $C = L (s^{L+1} - s) / (s - 1)$ is a normalization factor to ensure that the total number of edges remains $E$.

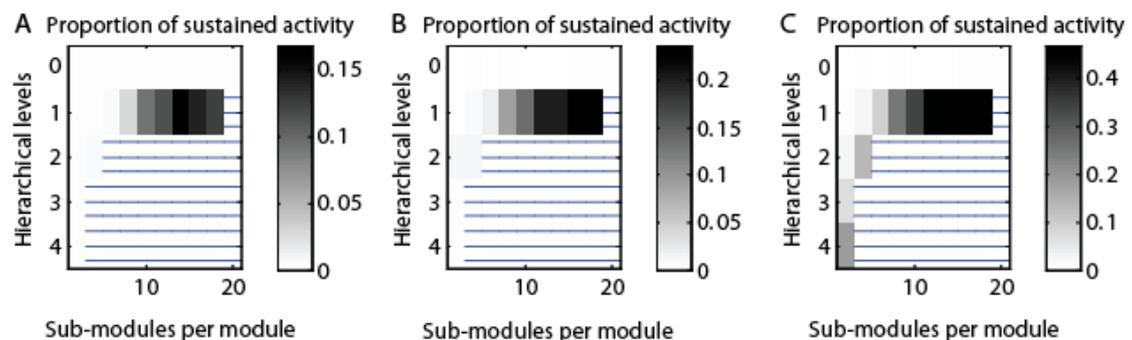

**Figure A1: Varying the number of edges per hierarchical level.** (**A**) Decreasing number of edges for higher hierarchical levels ($E_i \sim (2/3)^i$). (**B**) Number of edges independent from hierarchical level ($E_i$ = const.). (**C**) Increasing number of edges for higher hierarchical levels ($E_i \sim (3/2)^i$).

As shown in Figure A1, configurations with a high proportion of LSA for decreased as well as increased numbers of edges per hierarchical level remained comparable with the original calculation. The absolute level of sustained activity was lower in case (a) and higher in case (b) compared to the original settings. In addition, sustained activity cases also occurred for two or more hierarchical levels when the number of edges was increased (Figure A1C).

**Varying the parcellation for different hierarchical levels**
In the default settings, we used the same split-up of modules for each level; that means, if a module consisted of two sub-modules at the highest network level, this condition would be the same for all other levels of the network hierarchy as well. Here, we tested varying the parcellation into sub-modules depending on the hierarchical level. Again, we tested two cases: (a) creating fewer sub-modules for higher hierarchical levels ("decreasing parcellation") or (b) creating more sub-modules for higher hierarchical levels ("increasing parcellation"). The change followed a function where the number of parcellations, sub-modules per module, $m_i$ at level $i$ was given by $m_i = s^i m$ where $s$ is a scaling factor of 0.9 for decreased and 1.1 for increased parcellation. The parameter $m$ was the same as for the original calculation.

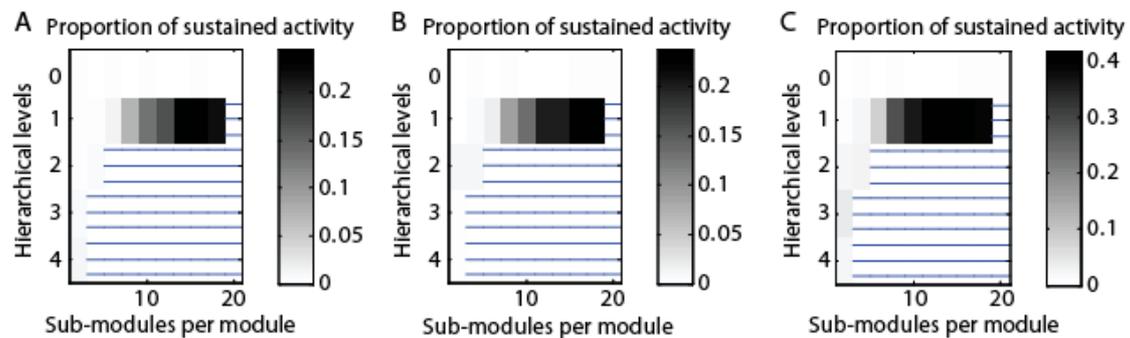

**Figure A2: Varying the parcellation (number of sub-modules per module) for hierarchical levels.** (**A**) Decreasing number of sub-modules $m_i$ for higher hierarchical levels ($m_i \sim 0.9^i$). (**B**) Parcellation into sub-modules independent from hierarchical level ($m_i = m$ = const.; see main text). (**C**) Increasing number of sub-modules $m_i$ for higher hierarchical levels ($m_i \sim 1.1^i$).

As can be seen from Figure A2, configurations with a high probability of sustained activity for decreased as well as increased number of sub-modules for each hierarchical level remained comparable with the original calculations. However, whereas the absolute proportions for decreased parcellation were similar, for increased parcellation the overall probabilities increased, extending the maximum proportion of sustained activity cases from 0.23 to 0.42. Therefore, the absolute parameter range of sustained activity was the same for case (a) and almost twice as high for case (b), compared to the original setting.

Note that both an increased parcellation and a larger number of edges led to a higher edge density of modules at the highest (fine-grained) level (maximum edge density, Figure 3B). This could explain why the ratio of sustained activity cases was higher for these configurations.

**Number of cases close to 50% activation threshold for classification as sustained**
In these simulations, we tested how close the final activity after 200 time steps came to the threshold used for classification as sustained activity case.

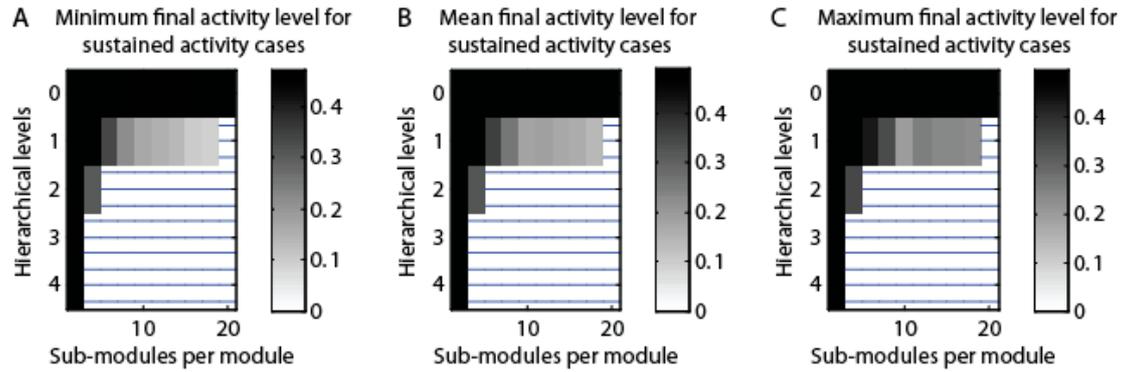

**Figure A3: Final activity for n=20 runs classified as sustained activity.** (**A**) Minimal final activity level. (**B**) Average final activity level. (**C**) Maximum final activity level.

As seen in Figure A3, levels close to the 50% threshold did occur. However, final activity levels were around 10-20% in all cases for which a high number of sustained activity cases was reported. This observation indicates that configurations with high proportions of sustained activity cases were not affected by the threshold. On the other hand, configurations with a high proportion of cases with complete spreading only had few cases of sustained activity. Due to higher activity levels for such configurations, sustained activity cases were close to the 50% threshold for classifying sustained activity (black regions in Figure A3).

**Topologies leading to expiring, limited sustained and completely spreading activity patterns**
The main simulations of this project investigated the proportion of LSA cases depending on hierarchical network organization. Here, we also considered the distribution of the other two possible simulation outcomes: activity dying out before the end of the simulation and final activity in more than 50% of the nodes, which was classified as complete spreading. Figure A4 shows the dependence of all three outcomes on hierarchical network organization.

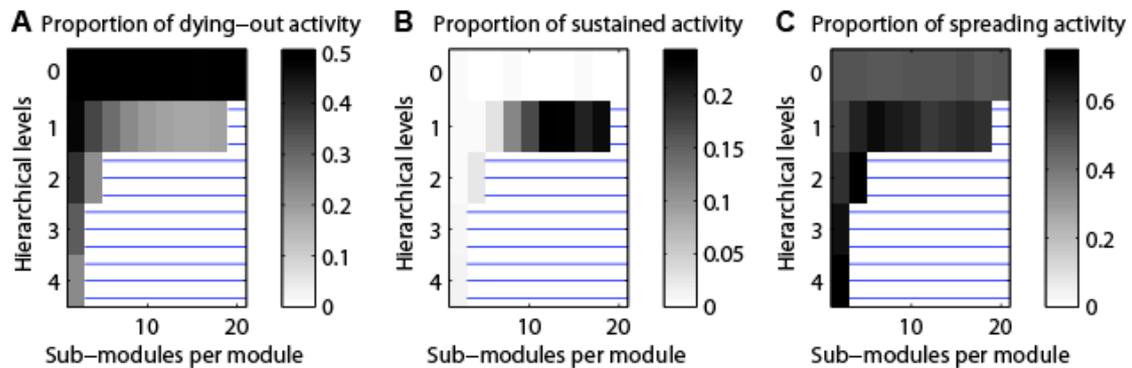

**Figure A4: Proportion of cases resulting in one of three scenarios of final activity level.** (**A**) Activity dying out. (**B**) Limited sustained activity. (**C**) Activity spreading through the network (above 50% activation threshold).

For configurations resulting in a small number of LSA cases (white regions in Figure A4B), both dying-out and complete spreading occurred in 50% of the cases (note the different gray level setting due to re-scaling). However, when more than one hierarchical level was present, complete spreading occurred more often than dying-out. For one hierarchical level, outcomes depended strongly on the number of sub-modules per module. The cases in which activity expired formed 50% of cases for two sub-modules, but decreased with the number of sub-modules. The maximum proportion of complete spreading, around 70% of the cases, occurred for 4-10 sub-modules per module. The maximum values for LSA occurred for 12-20 sub-modules per module.

**Varying the edge density**
In the main simulations, the edge density for a network with $N$=512 nodes and an average node degree $<k>$ of 50 was 0.098, that means, around 10%. Both parameters, edge density and average node degree are related: the edge density in a directed network is given by $d = E / (N(N-1))$ whereas the average node degree is $<k> = E / N$, meaning that $d = <k> / (N-1)$. How does variation of edge density for the same network size influence the range of LSA? To answer this question, we compared edge densities which were half (5%) of or twice (20%) of the value of the original calculations.

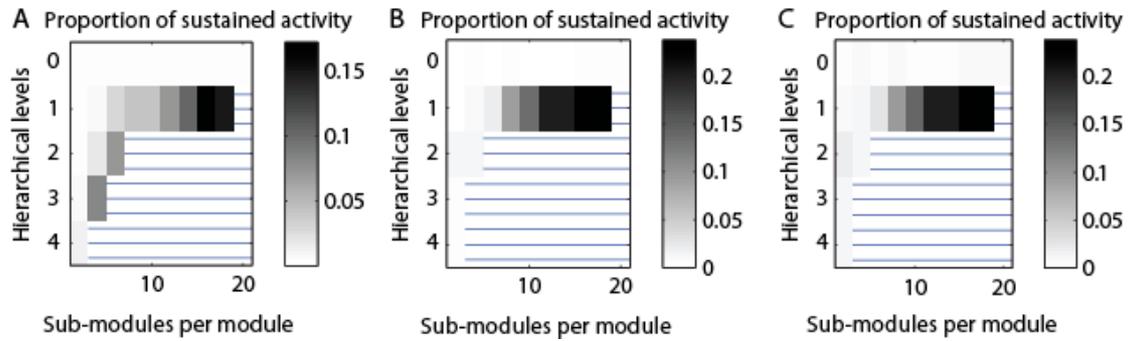

**Figure A5: Varying the edge density in a network with 512 nodes.** (**A**) Decreased edge density $d$=5%, average node degree $<k>$=25. (**B**) Original edge density $d$=10%, average node degree $<k>$=50. (**C**) Increased edge density $d$=20%, average node degree $<k>$=100.

As shown in Figure A5, the pattern of sustained activity remained comparable to the original calculations even when the edge density varied. The maximum level of LSA varied between 0.172 (for decreased edge density) to 0.238 (for increased edge density; original maximum level: 0.238). The relative distribution for one hierarchical level was similar across different edge densities. For two and three hierarchical levels, two additional configurations with high levels of sustained activity occurred for decreased edge density. These additional configurations were impossible for higher numbers of edges to be realized when edge densities were around 10% or above.